\documentclass[aps,twocolumn,longbibliography]{revtex4}
\usepackage{amsmath} 
\usepackage{graphicx}
\usepackage{subfig}
\usepackage{bbold}
\usepackage{float}
\usepackage[pdftex]{color}
\usepackage[colorlinks, linkcolor=blue, citecolor=blue, urlcolor=blue, breaklinks=true]{hyperref}
\usepackage{hyperref}

\begin{document}
\title{A superradiant two-level laser with intrinsic light force generated gain}
\author{Anna Bychek$^{1}$}
\author{Helmut Ritsch$^{1}$}
\affiliation{
$^1$Institut f\"ur Theoretische Physik, Universit\"at Innsbruck, Technikerstr. 21a, A-6020 Innsbruck, Austria}
\date{\today}

\begin{abstract}
The implementation of a superradiant laser as an active frequency standard is predicted to provide better short-term stability and robustness to thermal and mechanical fluctuations when compared to standard passive optical clocks. However, despite significant recent progress, the experimental realization of continuous wave superradiant lasing still remains an open challenge as it requires continuous loading, cooling, and pumping of active atoms within an optical resonator. Here we propose a new scenario for creating continuous gain by using optical forces acting on the states of a two-level atom via bichromatic coherent pumping of a cold atomic gas trapped inside a single-mode cavity. Analogous to atomic maser setups, tailored state-dependent forces are used to gather and concentrate excited-state atoms in regions of strong atom-cavity coupling while ground-state atoms are repelled. To facilitate numerical simulations of a sufficiently large atomic ensemble, we rely on a second-order cumulant expansion and describe the atomic motion in a semi-classical point-particle approximation subject to position-dependent light shifts which induce optical gradient forces along the cavity axis. We study minimal conditions on pump laser intensities and detunings required for collective superradiant emission. Balancing Doppler cooling and gain-induced heating we identify a parameter regime of a continuous narrow-band laser operation close to the bare atomic frequency.
\end{abstract}

\maketitle


\section{Introduction}

In view of establishing a new outstanding and robust optical time and frequency standard the quest to build a continuous superradiant laser operating on a very narrow atomic transition has been the subject of intense theoretical and experimental research in the past decade \cite{meiser09,bohnet12,maier14,bychek21,norcia18,debnath18,liu20,wu22,zhang22,kazakov22}. These studies are also fueled by the remarkable operating characteristics and relative simplicity of its microwave analogue, the hydrogen maser \cite{weiss96, bauch15}.
Recently pulsed superradiance has been experimentally observed using laser-cooled atomic ensembles \cite{norcia16science,laske19,tang21}. Some proof of principle setups based on magneto-optical trapping demonstrated quasi-continuous operation on kHz transitions \cite{norcia16, kristensen23}. The major remaining challenge is to achieve sufficient gain via continuous inversion on the relevant clock transition without significantly perturbing the atomic levels. 

One straightforward approach which is currently pursued is based on a continuous ultracold beam of excited atoms passing through an optical resonator \cite{bohnet12,  liu20, chen19, escudero21, chen22, tang22, cline22, takeuchi23}. 
In the past years, considerable progress has been made in this direction, yet the main challenge is to create a sufficient inverted atomic flux needed for collective superradiant emission.
As an alternative, some calculations suggest optimized multilevel pumping schemes, where a careful choice of laser powers and detunings minimizes the transition level shifts at reasonable pumping rates \cite{hotter22}. Still this needs many lasers, which have to be combined to pump and  cool the atomic ensemble simultaneously in order to keep the density of the active gas constant.

As in a standard micromaser, where population inversion is created by coherent pumping of atoms followed by magneto-mechanical separation of ground- and excited-state atoms in the mode volume, one could look for an analogous scheme for an optical setup. As the length scales are several orders of magnitude smaller at optical frequencies, the magnetic gradients for a sufficient state separation are very difficult to achieve and also potentially shift the clock transition in a detrimental way. However, one could make use of state-dependent optical forces, and thus excitation and separation could be done by suitably designed laser fields. 
At the same time, it has already been shown in previous experiments on BEC formation in Strontium that dressing lasers can create large enough optical level shifts in a dimple configuration to manipulate only a chosen sub-ensemble of the atoms with an extra laser without affecting the majority of the unperturbed atoms outside the dimple~\cite{chen22}.

Here, we suggest a new scenario for creating an intrinsic light force generated inversion mechanism. The idea is to combine the internal degrees of freedom of atoms with the motional ones to create the necessary inversion. 
We will study configurations, where pumping and gain occur in different spatial regions of the cavity by taking into account atomic motion and state-dependent forces resulting from a spatially dependent periodic drive of the transition.
After all, an inverted ensemble of only very weakly perturbed atoms can be created in regions of strong atom-cavity coupling. For sufficiently many atoms we show that this should lead to collective narrow-band lasing close to the bare atomic transition frequency. 

A detailed quantum description including the necessary number of atoms to achieve sufficient gain at low excitation powers goes beyond the available computational power to numerically simulate the coupled atom-field dynamics. Therefore, we have to resort to approximations and only limited atom numbers from which we are able to extract predictions for scaling towards larger ensembles. Hence we treat the atomic motion semi-classically and use a quantum description only for the internal atomic dynamics and the cavity field. As it has been observed for instance in Ref.~\cite{horak98}, a semi-classical description of motion shows a good agreement with the full quantum description, where the external degrees of freedom are quantized.
Still, we have to make use of a cumulant expansion approach~\cite{kubo62} for studying larger atom numbers.   

The paper is organized as follows. First, in Sec.~\ref{Light shifts} we introduce the spatial light shifts and optical forces present in the system. In Sec.~\ref{Coupled atom-field dynamics}, we present the system overview and calculate the coupled atom-field dynamics under the coherent laser drive. 
It is then shown in Sec.~\ref{Bichromatic laser drive}, that a bichromatic coherent drive can lead to a continuous narrow-band laser operation. We extend the model for an ensemble of atoms with light force induced inversion in Sec.~\ref{Collective dynamics}.
We start with the full quantum approach in Sec.~\ref{Full quantum} and proceed with the second-order cumulant expansion in Sec.~\ref{Second-order} in order to describe the collective atomic dynamics and spectrum of the light field in the cavity.

\section{Model definition}
\label{Model definition}

\subsection{Light shifts and forces}
\label{Light shifts}

A two-level atom coherently driven by a laser detuned from the atomic resonance frequency experiences energy light shifts \cite{dalibard85}. Under this drive the ground and excited states of the atom are no longer eigenstates of the system. The Hamiltonian of the atom ($\omega_a$) under the coherent laser drive ($\omega_\Omega$) in a rotating frame of the laser field can be written as ($\hbar = 1$) 

\begin{equation}
\label{Hamiltonian}
H_a = -\Delta_a \sigma^+ \sigma^- +\Omega (\sigma^+ +\sigma^-) = -\frac{\Delta_a}{2} \mathbb{1} -\frac{\Delta_a}{2}\sigma^z +\Omega \sigma^x,
\end{equation}
where $\Delta_a = \omega_\Omega - \omega_a$, $\Omega$ is the transition Rabi frequency, and $\sigma^x = \sigma^+ +\sigma^-$, $\sigma^z = \sigma^+\sigma^- - \sigma^-\sigma^+$ are the Pauli matrices. It is known that any 2x2 Hermitian matrix can be expressed in a unique way as a linear combination of the Pauli matrices
\begin{equation}
\label{Hamiltonian2}
H_{2\times 2} = h_0 \mathbb{1} +\vec{h}\vec{\sigma},
\end{equation}
with all coefficients being real numbers $h_0 = const$, $h_1 = h\sin{\Theta}\cos{\phi}$, $h_2 = h\sin{\Theta}\sin{\phi}$, $h_3 = h\cos{\Theta}$, where $h=|\vec{h}|=\sqrt{h_1^2 +h_2^2 +h_3^2}$. It is easy to show that the eigenvalues of this matrix are
\begin{equation}
\label{eigenvals0}
E_{\pm} = h_0 \pm h,
\end{equation}
and the corresponding eigenvectors can be expressed as an effective rotation of the uncoupled states,
\begin{equation}
\begin{aligned}
\label{eigenvecs0}
|+\rangle &= \sin{\frac{\Theta}{2}}\exp{(\frac{i\phi}{2})}|g\rangle + \cos{\frac{\Theta}{2}}\exp{(-\frac{i\phi}{2})}|e\rangle \cr
|-\rangle &= \cos{\frac{\Theta}{2}}\exp{(\frac{i\phi}{2})}|g\rangle - \sin{\frac{\Theta}{2}}\exp{(-\frac{i\phi}{2})}|e\rangle.
\end{aligned}
\end{equation}
Therefore, the eigenvalues of the Hamiltonian~(\ref{Hamiltonian}) can be written as
\begin{equation}
\label{eigenvals}
E_{\pm} = -\frac{\Delta_a}{2} \pm \sqrt{\Omega^2+\Delta_a^2/4},
\end{equation}
with the corresponding eigenstates known as the dressed states,
\begin{equation}
\begin{aligned}
\label{eigenvecs}
|+\rangle &= \sin{\frac{\Theta}{2}}|g\rangle + \cos{\frac{\Theta}{2}}|e\rangle \cr
|-\rangle &= \cos{\frac{\Theta}{2}}|g\rangle - \sin{\frac{\Theta}{2}}|e\rangle,
\end{aligned}
\end{equation}
where $tg \:\Theta = -\frac{2\Omega}{\Delta_a}$. 

\begin{figure*}[!t]
    \centering
    \includegraphics[width=0.99\textwidth]{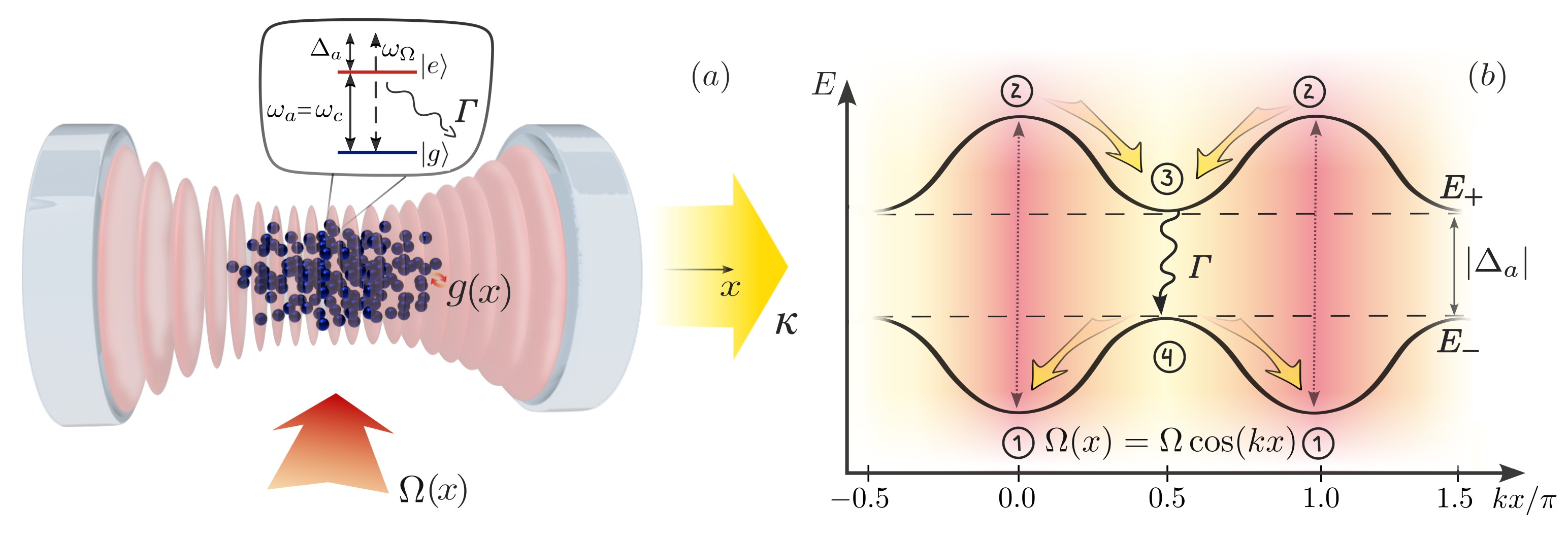}
    \caption{Schematics of the system. (a) Two-level atoms ($\omega_a$) inside a single-mode cavity ($\omega_c$) are coherently driven by a laser ($\omega_\Omega$) with the Rabi frequency $\Omega(x)$, where $\Gamma$ is the single-atom spontaneous emission rate, $g(x)$ is the atom-cavity coupling, and $\kappa$ is the cavity loss rate. 
    (b) Eigenvalues of the Hamiltonian given in Eq.~(\ref{Hamiltonian}) under a spatially dependent coherent laser drive $\Omega(x) = \Omega \cos(kx)$ with the negative detuning $\Delta_a = \omega_\Omega - \omega_a$, where yellow arrows show a dipole force acting on the states of the two-level atom as described in the main text.}
    \label{Fig1}
\end{figure*}

Let us now consider a two-level atom moving in one dimension along the axis of a linear cavity, as schematically presented in figure~\ref{Fig1}(a). When the atom is illuminated by a laser whose Rabi frequency has a spatial periodic distribution ${\Omega(x) = \Omega \cos(kx)}$ formed by a standing wave with the wavelength $\lambda = 2\pi/k$ the energy shifts can be plotted as shown in figure~\ref{Fig1}(b). This creates a mean dipole force ${\langle F \rangle = -\langle \nabla H \rangle}$ acting on the atom, which has the opposite sign for $|+\rangle$ and $|-\rangle$ states. In this regard one could think of a population inversion scheme shown in figure~\ref{Fig1}(b). An atom located at position (1) with some non-zero initial velocity is pumped by the laser into state (2) and experiences the dipole force as it continues to move. This force pushes the atom to position (3) where there is no force acting on the atom. If this process occurs at a faster rate than the lifetime $\tau \sim \Gamma^{-1}$ of the excited state, then the atom emits a photon with the frequency close to the bare atomic transition frequency and undergoes the transition into state (4) where it is dragged by the light force left or right to position (1) and the process repeats itself. 

Thus, such a scheme could be used to spatially separate the region of pumping from the lasing in the system. In other words, we create the population inversion in the specific regions of the cavity, those regions where we would like to have a maximal coupling to the cavity.

\subsection{Semi-classical master equation for the coupled atom-field dynamics}
\label{Coupled atom-field dynamics}

A stable laser operation requires a continuous inversion mechanism to keep the population inversion on a lasing transition. In the previous section, we introduced the scenario for creating effective inversion on a lasing transition using light forces.
While in principle many possible geometries to achieve this purpose can be investigated, we will restrict ourselves here to a simple generic case, where the underlying mechanisms can be studied in detail. Hence we consider a 1D motion in a single-mode Fabry-Perot cavity with a sine wave laser mode and apply a standing cosine wave pump field. Ground-state atoms will be trapped and cooled close to the antinodes of the cosine mode. Atoms excited to the upper level at these points are pushed towards the nodes of the cosine mode, where they maximally couple to the cavity sine mode.
Let us identify a parameter regime, which leads to the desired local inversion and gain.  In other words, one needs to find a regime of stable narrow-band lasing at the bare atomic frequency, with a linewidth that is much smaller than the cavity linewidth and that is well distinguished from the other light sources present in the cavity. This requires several conditions to be ensured: 

$\bullet$ $\kappa > \Gamma$ - the system is in the bad-cavity regime, which provides the low intracavity photon number operation and thus reduced sensitivity to cavity noise;

\vspace{0.2cm}
$\bullet$ $\Delta_a<0$, and $|\Delta_a| > \Omega$ - the driving laser is far red-detuned from the atomic transition frequency to minimize the amount of coherently scattered photons from the drive into the cavity;

\vspace{0.2cm}
$\bullet$ $2\sqrt{\Omega^2+\Delta_a^2/4} > \kappa$ - the maximal light shift in equation~(\ref{eigenvals}) is larger than the cavity linewidth to spatially separate the region of pumping (1-2) from the lasing (3-4) in figure~\ref{Fig1}(b).

The Hamiltonian of the two-level atom subjected to a coherent drive inside a single-mode optical cavity can be described by the Jaynes–Cummings model in the rotating frame of the drive
\begin{equation}
    \label{Hamiltonian_0}
H = -\Delta_a \sigma^+\sigma^- -\Delta_c a^{\dagger}a + g(x)(a^{\dagger}\sigma^- +a\sigma^+) +\Omega(x) (\sigma^+ +\sigma^-),
\end{equation}
where $\Delta_c = \omega_\Omega - \omega_c$ is the laser detuning from the cavity mode, $g(x)=g \sin(k x)$ and $\Omega(x)=\Omega \cos(kx)$ are the atom-cavity interaction strength and Rabi frequency, respectively, which are functions of the atomic position along the cavity mode with the wave number $k$.
The quantum dynamics of the open atom-cavity system can be described by the master equation for the system density matrix $\rho$ in the Lindblad form
\begin{equation}
    \label{master equation}
\dot{\rho} = -i[H, \rho] + \mathcal{L}_\kappa[\rho] + \mathcal{L}_\Gamma[\rho],
\end{equation}
where the loss of photons through the cavity mirrors and individual atomic decay are given by
\begin{equation}
\begin{aligned}
\label{eqs_Lindblad}
\mathcal{L}_\kappa[\rho] &= \frac{\kappa}{2}(2a\rho a^\dagger - a^\dagger a \rho - \rho a^\dagger a) \cr
\mathcal{L}_\Gamma[\rho] &= \frac{\Gamma}{2} (2\sigma^- \rho \sigma^+ -\sigma^+ \sigma^- \rho -\rho \sigma^+ \sigma^-),
\end{aligned}
\end{equation}
with the cavity loss rate $\kappa$ and single-atom spontaneous emission rate $\Gamma$, respectively. In order to approximate the atomic motion and light forces acting on the atom, we include the semi-classical equations of motion in the system description, 
\begin{equation}
\begin{aligned}
\label{eqs_motion}
\dot{\langle x \rangle} &= \langle p \rangle /m = 2\omega_r \langle p \rangle / k_a^2 \cr
\dot{\langle p \rangle} &= -\langle \nabla H \rangle,
\end{aligned}
\end{equation}
where ${\omega_r = k_a^2/(2m)}$ is the atomic recoil frequency given by the atomic mass and wave number of the atomic transition. Here we neglect the effects of momentum diffusion arising from spontaneous emission. One could account for these effects by going beyond the mean-field approach in calculating the position and momentum, which would substantially increase the amount of equations in our simulations. At the same time, in the ultimate case of a superradiant laser, the emission rate $\Gamma$ is usually very small compared to other parameters, such that spontaneous emission recoil should only lead to minor corrections in the atomic dynamics.

First of all, we would like to calculate the coupled atom-photon dynamics in the cavity. For a non-moving atom under a coherent drive the solution is known as the damped Rabi oscillations eventually leading to population of the excited state by no more than fifty percent \cite{fox06}, i.e.~no inversion. However, due to atomic motion and forces acting differently on the states of the atom, the local  population inversion can become positive in certain positions.

\begin{figure*}[!t]
    \centering
    \includegraphics[width=0.95\textwidth]{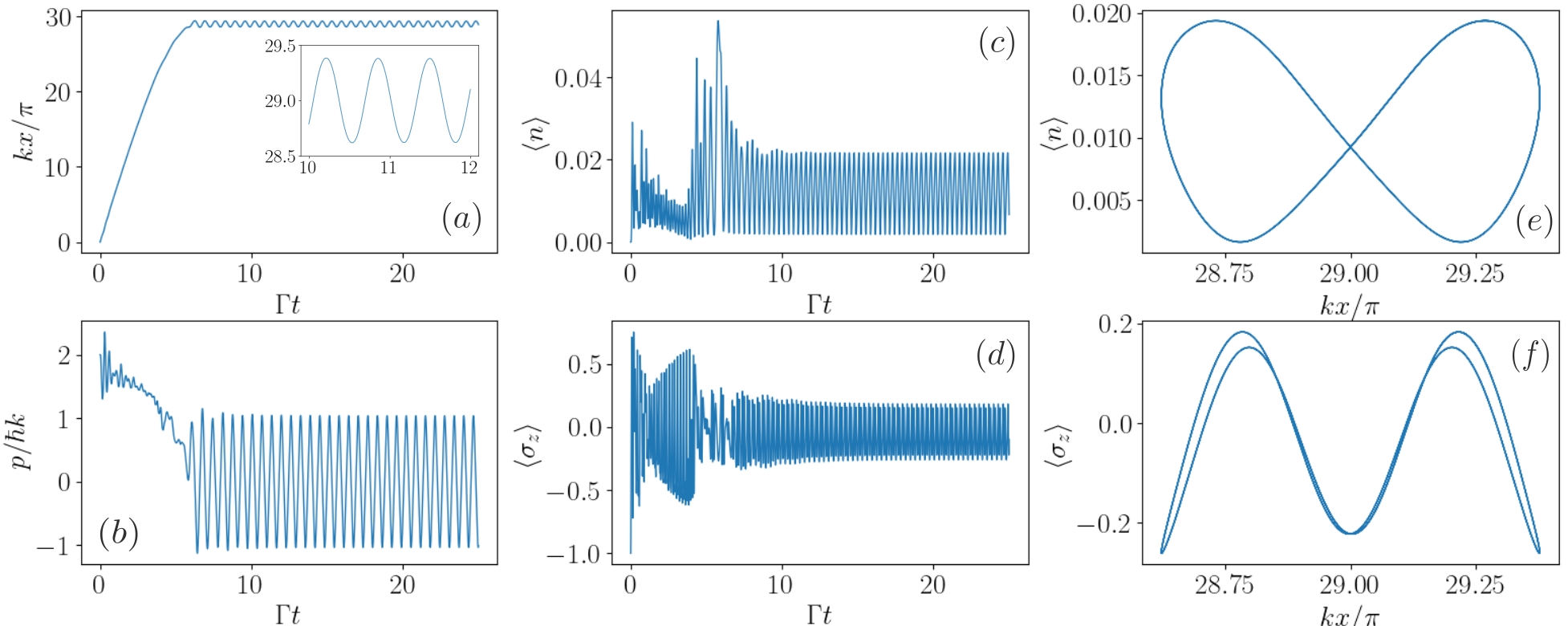}
    \caption{Atomic dynamics and lasing under the coherent drive with $|\Delta_a|<\Omega$. Parameters: $\kappa = 20\Gamma$, $g = 4\Gamma$, $\Omega = 30\Gamma$, $\Delta_a = -10\Gamma$, $\omega_r = 6\Gamma$.}
    \label{Fig2}
\end{figure*}

\begin{figure}[b!]
    \centering
    \includegraphics[width=\columnwidth]{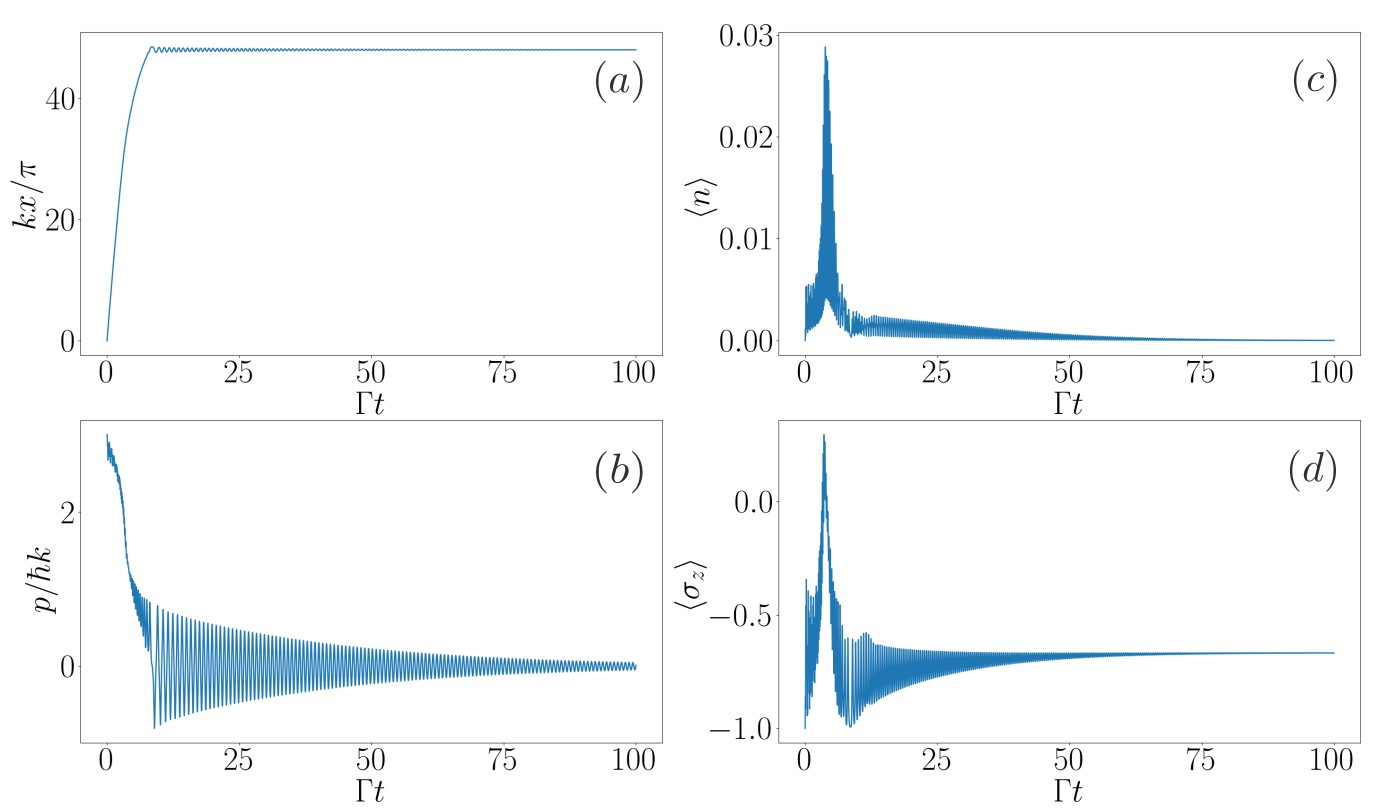}
    \caption{Cooling and trapping of an atom in the cavity under the coherent drive with $|\Delta_a|>\Omega$. Parameters: $\kappa = 20\Gamma$, $g = 4\Gamma$, $\Omega = 10\Gamma$, $\Delta_a = -20\Gamma$, $\omega_r = 6\Gamma$.}
    \label{Fig3}
\end{figure}

For the remainder of this work, we set the cavity mode to be on resonance with the bare atomic transition frequency and select a recoil frequency value that facilitates rapid cooling for an adequate representation of the results. We consider a linear cavity with a cosine wave pump $\Omega(x)= \Omega \cos(kx)$ and a sine wave cavity mode with the coupling strength $g(x) = g \sin(kx)$, where $k = k_a$ is the cavity mode wave number.  Since we are in the parameter regime where $\kappa > \Gamma$ and $g < \kappa$, the Hilbert space of the photon field can be truncated at low photon numbers.
Figures~\ref{Fig2}-\ref{Fig3} show the atomic dynamics and lasing under the coherent drive for two different cases of $|\Delta_a|<\Omega$ and $|\Delta_a|>\Omega$. The position and momentum of the atom are given in units of $\pi/k$ and $\hbar k$, respectively, in figures~\ref{Fig2}(a)-(b).  The mean photon number in the cavity and atomic inversion are shown as a function of time in figures~\ref{Fig2}(c)-(d) and a function of position in figures~\ref{Fig2}(e)-(f).

In the case of $|\Delta_a|<\Omega$, we have found the parameter regime, where atomic cooling is balanced by heating from the driving laser, see figures~\ref{Fig2}(a)-(b). Starting from a given initial momentum, the atom experiences laser cooling until it reaches a quasi-stationary state. As seen in figure~\ref{Fig2}(a), the atom oscillates between a particular node and the neighboring antinodes of the driving field while the atomic inversion $\langle \sigma_z \rangle = \langle \sigma^+ \sigma^- \rangle - \langle \sigma^- \sigma^+ \rangle$ in figure~\ref{Fig2}(f) becomes positive towards the points of the maximal coupling strength. Followed by photon emission and a maximum in average photon number in figure~\ref{Fig2}(e) this dynamics is close to the ideal scenario depicted in figure~\ref{Fig1}(b). The results demonstrate that in principle the above scenario may take place, unfortunately this can not be used as a good pumping scheme due to the fact that mostly scattered photons from the drive will dominate the cavity output spectrum. 

On the other side, after studying the case of $|\Delta_a|>\Omega$, we have observed that the drive from a single laser is never sufficient enough to create the desired population inversion. As presented in figure~\ref{Fig3}, in this case the applied far-detuned drive results in strong cooling for the atom and there is no inversion on the lasing transition at any point. Therefore, one may think of the idea of adding an extra laser drive to populate the excited state, with the parameters particularly chosen and optimized to ensure the above conditions in the cavity.

\subsection{Two-level dynamics with a bichromatic coherent drive}
\label{Bichromatic laser drive}

\begin{figure*}[t!]
    \centering
    \includegraphics[width=0.97\textwidth]{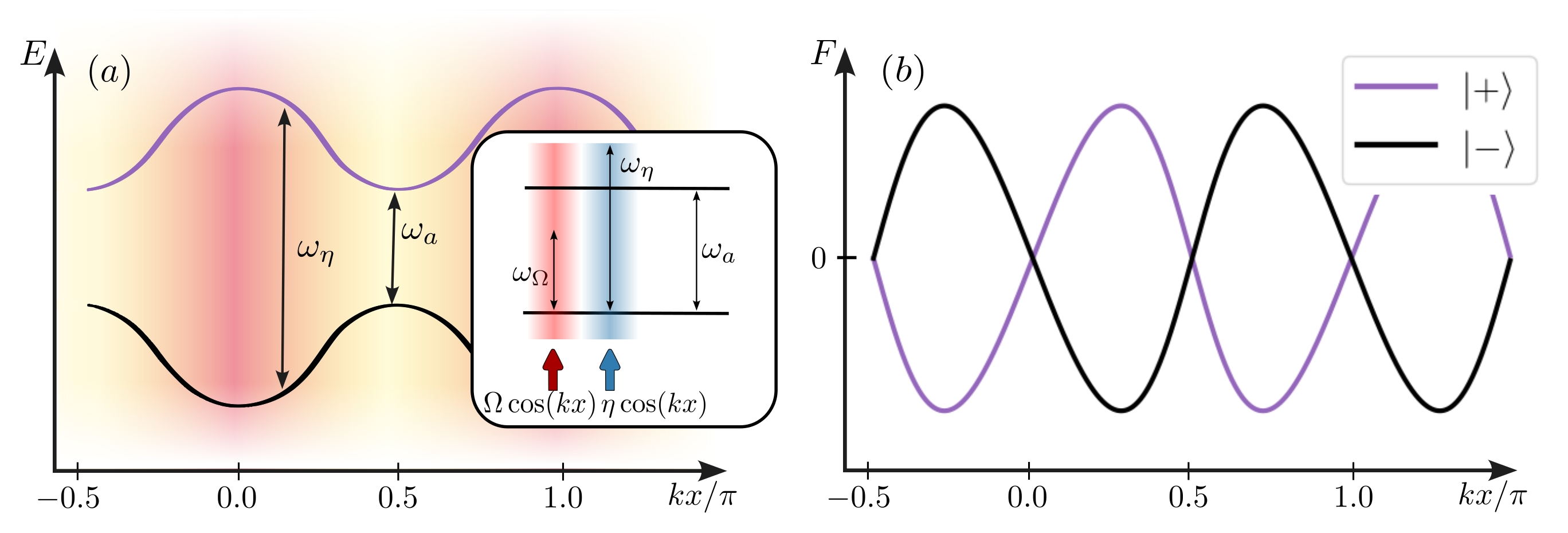}
    \caption{(a) The schematics of an atom under the bichromatic coherent drive. (b)~Dipole force acting on the dressed states of the atom.}
    \label{Fig4}
\end{figure*}
\begin{figure*}[t!]
    \centering
    \includegraphics[width=0.98\textwidth]{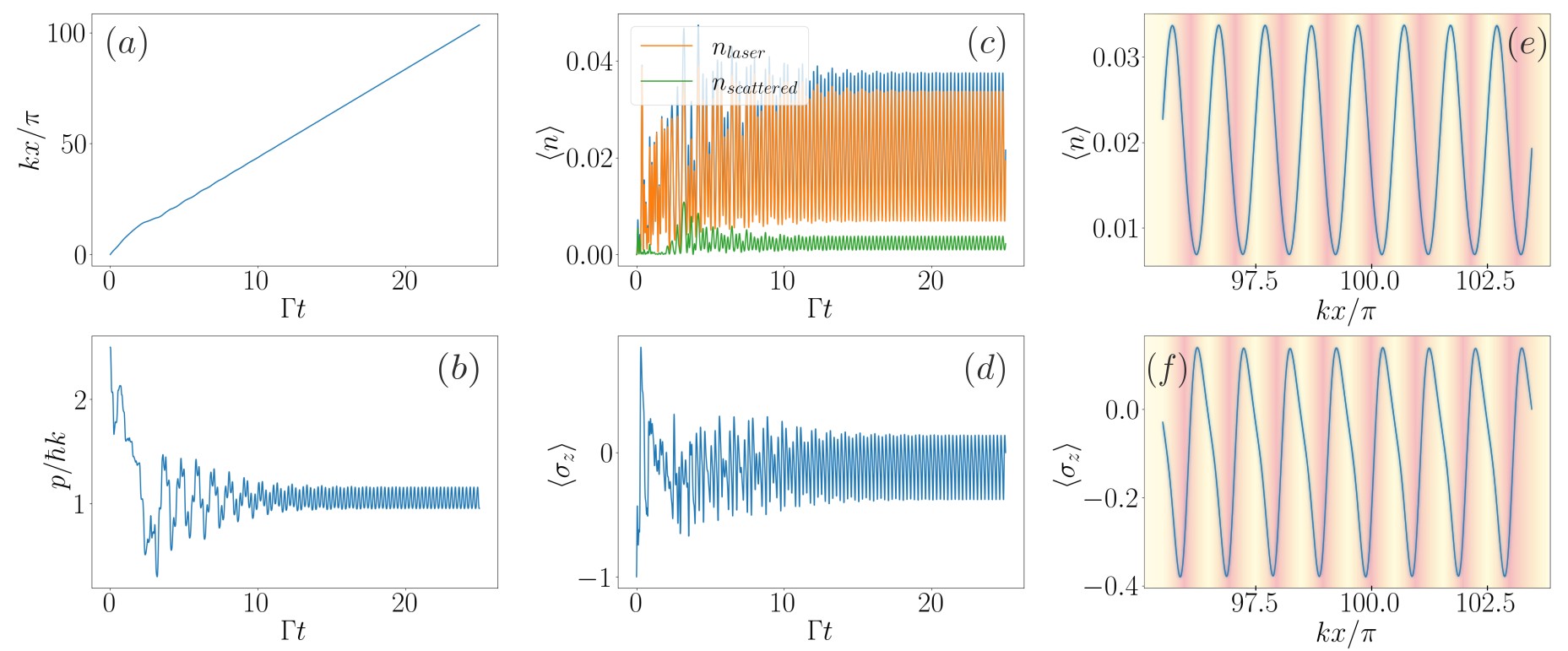}
    \caption{Atomic dynamics and lasing under the bichromatic coherent drive. Parameters: $\kappa = 20\Gamma$, $g = 4\Gamma$, $\Omega = 10\Gamma$, $\Delta_a = -20\Gamma$, $\eta = 8\Gamma$, $\Delta_\eta = -25\Gamma$, $\omega_r = 6\Gamma$.}
    \label{Fig5}
\end{figure*}

In the previous section, we have indicated the parameter regime of our interests in the context of the superradiant laser.  We observed that the driving from a single laser is not sufficient to create the desired population inversion on the lasing transition. In this section, the idea is to use this laser exclusively to create the spatial light shifts, as depicted in figure~\ref{Fig4}(a). In order to populate the excited state we introduce the second laser drive $\eta(x) = \eta \cos(kx)$ into the system. The frequency of the second laser drive is now tuned to the resonance with the dressed states given by equations~(\ref{eigenvals})-(\ref{eigenvecs}), however not at their maximal light shifts, but at the points where the dipole force acting on the atom is close to its maxima, see figure~\ref{Fig4}(b). This allows the excited atom to reach the lasing position more efficiently since there is strong acceleration from the force. We expect that a combination of these laser drives acting together can lead to collective narrow-band emission, provided that optimal driving intensities and detunings are found. However, the master equation becomes significantly more difficult to solve as the second laser drive does not allow to eliminate the time dependence in the Hamiltonian. Thus the Hamiltonian of the system in the rotating frame of the first laser can be written as
\begin{equation}
\label{Hamiltonian_2lasers}
H_2 = H + \eta \cos(kx) (\sigma^+ e^{i\Delta_{\eta} t} +\sigma^- e^{-i\Delta_{\eta} t}),
\end{equation}
where $\Delta_\eta=\omega_\Omega - \omega_\eta$.
Note, that if the laser drive is not strong enough it will not be able to create population inversion. On the other hand, a strong laser drive will produce both a lot of coherently scattered photons and strong heating in the system. In order to find the dynamics of the system we solve the master equation~(\ref{master equation}) with the Hamiltonian~(\ref{Hamiltonian_2lasers}). It is then the Rabi frequency and detuning of the second laser which have to be scanned and mutually adjusted.

\begin{figure*}[t!]
    \centering
    \includegraphics[width=0.99\textwidth]{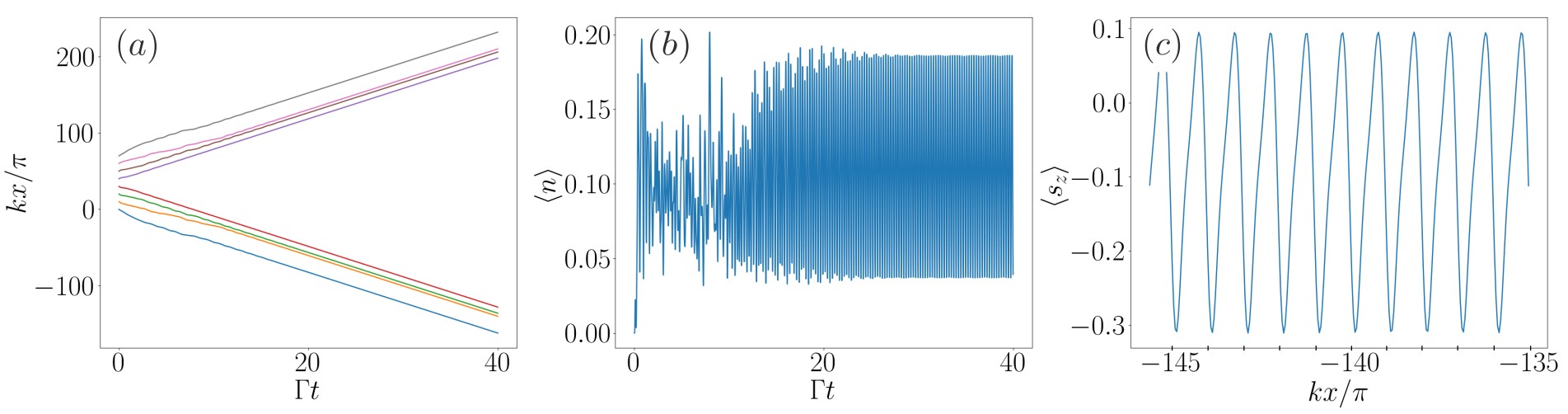}
    \caption{(a)~Dynamics of $N=8$ atoms under the bichromatic coherent drive. (b)~Mean photon number in the cavity depending on time (in units of $\Gamma$). (c) The position dependence of the atomic inversion during the stable final stage of the time evolution. The initial distribution of momenta $|p^0_m| \in \{ 2, ..., 2.5 \}\hbar k$ and positions $x^0_m = m\pi/ k$ for $m=1..N$. The parameters are the same as in figure~\ref{Fig5}.}
    \label{Fig6}
\end{figure*}
\begin{figure*}[t!]
    \centering
    \includegraphics[width=0.99\textwidth]{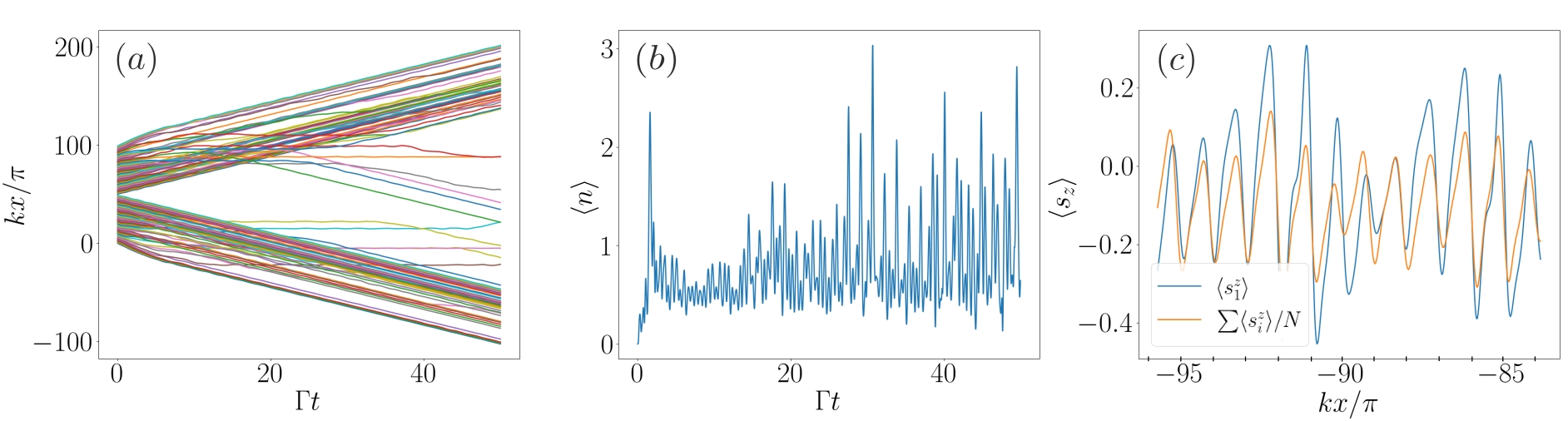}
    \caption{Atomic dynamics and lasing in the ensemble of $N=100$ atoms under the bichromatic coherent drive. The parameters are the same as in figure~\ref{Fig5}.}
    \label{Fig7}
\end{figure*}

Figure~\ref{Fig5} shows the atomic dynamics and lasing under the bichromatic coherent drive in the regime, where the light from the second drive pushes the atom to move along the cavity axis such that the atomic momentum reaches its quasi-stationary state, see figures~\ref{Fig5}(a)-(b). This results in a linear motion of the atom,  where it is being pumped at the maximal driving intensity and continues to undergo the inversion scheme depicted in figure~\ref{Fig1}(b). As the atom moves, the atomic inversion becomes positive in the vicinity of the unperturbed bare atomic transition frequency, which is followed by the photon emission, as can be seen in figures~\ref{Fig5}(d)-(f). Figure~\ref{Fig5}(c) shows the mean photon number in the cavity (blue), which can be split into the laser part (orange) and coherently scattered part (green). The coherently scattered field amplitude is phase dependent as it comes from the coherent laser drive and its intensity can be calculated as $|\langle a \rangle|^2$, while the incoherently scattered field from inverted atoms is phase independent. To obtain the laser part we subtract the coherently scattered part from the total photon number.
Furthermore, one can see a stable laser operation with the oscillations of the photon number around the mean value. This example demonstrates how the specially tuned bichromatic coherent drive can lead to continuous lasing with intrinsic light force induced inversion.

\section{Collective dynamics with light force induced inversion}
\label{Collective dynamics}

\subsection{Full quantum approach}
\label{Full quantum}

An extension of the full quantum model to an ensemble of atoms remains feasible only for a small atom number due to the exponential growth of the Hilbert space. Here we present the results for the case of $N=8$, where we truncate the Hilbert space of the field in the cavity at low photon numbers.  We choose the initial atomic momenta to be randomly distributed around a selected velocity as $|p^0_m| \in \{ p_0 - \epsilon, ..., p_0+\epsilon \}$,  which is of the order of several $\hbar k$ and $x^0_m = m\pi/ k$ for $m=1..N$, see figure~\ref{Fig6}. 
In addition, even when initial atomic momenta are one or even two orders of magnitude larger than $\hbar k$ one can observe the laser cooling process. As such atoms do not contribute to the desired lasing until they get cold enough, it substantially reduces the efficiency of the scheme. On the other hand, atoms with the initial velocity close to zero immediately become trapped in the potential minima and do not contribute to the dynamics as well. Therefore, the ideal initial conditions would be to primarily reduce the temperature of atoms and perform a velocity selection. As the atoms move one can see the stabilization of photon emission and increase in the mean photon number. Figure~\ref{Fig6}(c) shows how the atomic inversion changes with the position similar to the case of a single atom in figure~\ref{Fig5}.

\subsection{Second-order cumulant expansion}
\label{Second-order}

Next, we would like to extend our model to an ensemble of $N \gg 1$ atoms subjected to the bichromatic coherent drive, as described in the previous section. As each atom in the ensemble behaves differently depending on its initial position and momentum, the numerical solution of the master equation~(\ref{master equation}) with the Hamiltonian given in Eq.~(\ref{Hamiltonian_2lasers}) becomes challenging. In order to describe a large ensemble of atoms we make use of the second-order cumulant expansion \cite{kubo62} to write down the closed set of the Heisenberg equations for the system operators \cite{bychek21}:

\begin{widetext}
\begin{equation}
\begin{aligned}
\label{Heisenberg_eqs}
&\frac{d}{dt} \langle a \rangle =  -(\kappa/2-i\Delta_c) \langle a \rangle -ig\sum_j N_j \sin(x_j)\langle \sigma^-_j \rangle \cr
&\frac{d}{dt} \langle \sigma^-_m \rangle = -(\Gamma/2-i\Delta_{am})\langle \sigma^-_m \rangle +ig \sin(x_m) \langle a \rangle (2\langle \sigma^+_m \sigma^-_m \rangle -1) +i(\Omega+\eta e^{i\Delta_{\eta}t})\cos(x_m)(2\langle \sigma^+_m \sigma^-_m \rangle-1)\\
&\frac{d}{dt} \langle a^\dagger a \rangle = -\kappa \langle a^\dagger a \rangle + ig\sum_j N_j \sin(x_j) (\langle a \sigma^+_j \rangle - \langle a^\dagger \sigma^-_j \rangle) \\
&\frac{d}{dt} \langle a \sigma^+_m \rangle = -(\kappa/2+\Gamma/2+i\Delta_{am}-i\Delta_c)\langle a \sigma^+_m \rangle + ig\sin(x_m)(\langle a^\dagger a \rangle-2\langle a^\dagger a \rangle \langle \sigma^+_m \sigma^-_m \rangle-\langle \sigma^+_m \sigma^-_m \rangle)+... \\
&\quad ...-ig \sum_{j;m\neq j}\sin(x_j)\langle \sigma^+_m \sigma^-_j \rangle -i(\Omega+\eta e^{-i \Delta_{\eta}t})\cos(x_m)\langle a \rangle(2\langle \sigma^+_m \sigma^-_m \rangle-1) \\
&\frac{d}{dt}  \langle \sigma^+_m \sigma^-_m \rangle = -\Gamma \langle \sigma^+_m \sigma^-_m \rangle -ig\sin(x_m)(\langle a \sigma^+_m \rangle - \langle a^\dagger \sigma^-_m \rangle) +... \\
&\quad ...-i(\Omega+\eta e^{i \Delta_{\eta}t}))\cos(x_m)\langle \sigma^+_m \rangle +i(\Omega+\eta e^{-i \Delta_{\eta}t}))\cos(x_m)\langle \sigma^-_m \rangle \\
&\frac{d}{dt}  \langle \sigma^+_m \sigma^-_j \rangle = -\Gamma \langle \sigma^+_m \sigma^-_j \rangle -ig_m\sin(x_m)\langle a^\dagger \sigma^-_j \rangle (2\langle \sigma^+_m \sigma^-_m \rangle -1) +ig_j\sin(x_j)\langle a \sigma^+_m \rangle (2\langle \sigma^+_j \sigma^-_j \rangle -1) +... \\
&\quad ...+i(\Omega+\eta e^{i \Delta_{\eta}t}))\cos(x_j)\langle \sigma^+_m \rangle (2 \langle \sigma^+_j \sigma^-_j \rangle -1) -i(\Omega+\eta e^{-i \Delta_{\eta}t}))\cos(x_m)\langle \sigma^-_j \rangle (2 \langle \sigma^+_m \sigma^-_m \rangle -1) \\
&\frac{d}{dt}  \langle x_m \rangle = 2\omega_r \langle p_m \rangle / k^2  \\
&\frac{d}{dt}  \langle p_m \rangle = -2g\cos(x_m) \Re \{\langle a \sigma^+_m \rangle \} +2\Omega\sin(x_m) \Re \{\langle \sigma^+_m \rangle \} +\eta \sin(x_m)( \langle \sigma^+_m \rangle e^{i\Delta_{\eta}t} +\langle \sigma^-_m \rangle e^{-i\Delta_{\eta}t}),
\end{aligned}
\end{equation}
\end{widetext}
\begin{figure*}[t!]
    \centering
    \includegraphics[width=0.8\textwidth]{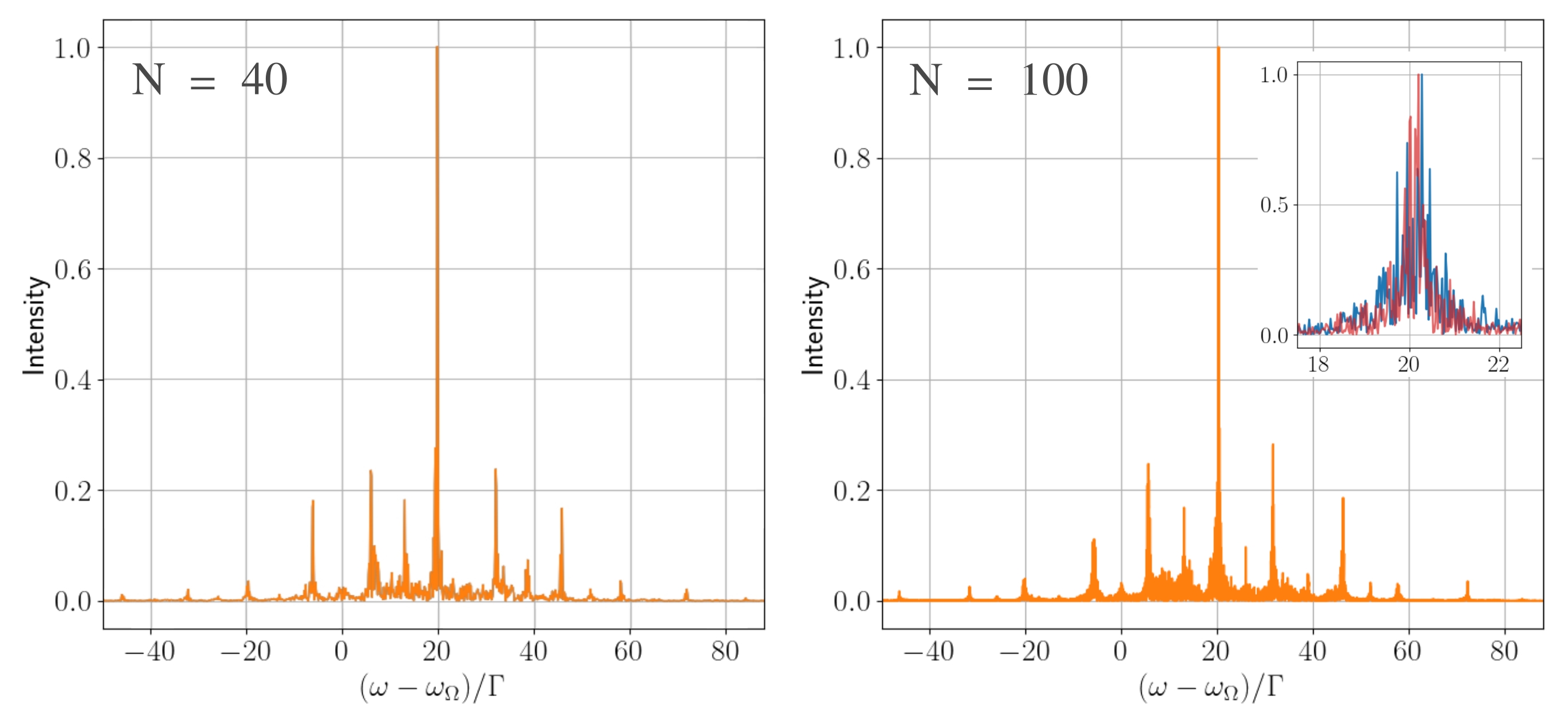}
    \caption{Spectrum of the cavity light for the ensemble of $N=40$ (left panel) and $N=100$ atoms (right panel) presented in figure~\ref{Fig7}. The emission intensity is normalized and the inset shows a zoom-in of the central peak profile (blue line). The red line in the inset shows the spectrum under cavity dephasing with the rate $\xi=10\Gamma$.}
    \label{Fig8}
\end{figure*}
where $m=1..N$ and $\Re$ is used to denote the real part of an expectation value of an operator. Figure~\ref{Fig7} shows the solution of equations~(\ref{Heisenberg_eqs}) for~$N=100$ atoms with the same distribution of initial positions and momenta as used in the full model considered in the previous section. The resulting dynamics becomes much more complicated to describe, but one can see a similar behavior with the case of a single atom. Although a small part of atoms gets cooled down and does not contribute to the lasing process, the majority of atoms display the lasing we are interested in. In figure~\ref{Fig7}(b) one can see continuous lasing from the atoms reaching the order of one photon on average in the cavity. We expect fluctuations in the photon number to be mitigated in the case of significantly larger atomic ensembles, where each atom only weakly contributes to the emission process. However, due to the number of equations growing as $\mathcal{O}(N^2)$ we are limited to a system of a few hundred atoms.

Let us calculate the spectrum of the light in the cavity. According to the Wiener–Khinchin theorem \cite{puri01} the spectrum can be found as a Fourier transform of the first-order correlation function $g^{(1)}(\tau) = \langle a^{\dagger}(t_0+\tau)a(t_0) \rangle$,
\begin{equation}
\label{WK_theorem}
S(\omega) = 2 \Re \left\{\int_{0}^{\infty} d\tau e^{-i\omega \tau} g^{(1)} (\tau) \right\}.
\end{equation} 
Here, we use the quantum regression theorem \cite{carmichael00} to write down a set of differential equations for the first-order correlation function, where $t_0$ is normally given by the time the system reaches its steady state.
However, since in our case the dynamics does not have a steady state, we have to include these equations in the full system of equations~(\ref{Heisenberg_eqs}) and average it over a set of equidistant initial conditions ${g^1(0) = \langle a^{\dagger}a\rangle (t_0)|_{t_0=t_{end}}}$ chosen from the final stage of the dynamics.

After the averaging process one can see the resulting spectrum, as presented in figure~\ref{Fig8} for $N=40$ and $N=100$ atoms.
The spectrum averages quite well already after several averaging steps and reveals the main spectral peak coming from the atoms at the frequency close to the bare atomic transition, which linewidth is below the natural linewidth of the atomic transition $\Gamma$. The inset shows a zoom-in of the central peak profile which broadening can be attributed to the emission from different atoms at slightly different frequencies (blue line). The red line shows that the resulting spectrum is robust to cavity fluctuations in the presence of moderate cavity noise  with the dephasing rate $\xi=10\Gamma$.

Additionally to the central peak, there are numerous sidebands located left and right from the atomic transition frequency. They appear to be independent of the number of atoms, coupling constant, and Rabi frequencies of the lasers. We associate them with the motion of atoms with a constant velocity, which is observed in figure~\ref{Fig7}(a). One can calculate the frequency of these motional sidebands as 
\begin{equation}
\label{motional_SB}
\omega_{SB} = \omega_a \pm \omega_{\pm} = \omega_a \pm 2\pi \frac{v}{\lambda},
\end{equation} 
where $v/\lambda = \omega_r p/(\pi k)$. Since the atomic motion is linear one can write
\begin{equation}
\label{motional_SB_2}
\omega_{\pm} = \pm \frac{2 \omega_r \langle p \rangle^{st}}{k}.
\end{equation} 
Substituting the real parameters used in figure~\ref{Fig8} into equations~(\ref{motional_SB_2}) we calculate the frequencies $w_{\pm}$, which agree well with the central frequencies of the sidebands observed in the spectrum.

\section{Conclusions and outlook}
\label{Conclusions}

We have studied population inversion and gain within an optical cavity in a cold ensemble of coherently driven two-level atoms with an intrinsic light force generated inversion mechanism. Using numerical simulations of the coupled atom-field dynamics we have found the operating conditions producing continuous narrow-band emission close to the unperturbed atomic line. In the limit of low photon number operation, the central frequency is largely insensitive to cavity fluctuations. Note, that the driving field has to be far-detuned from the atomic resonance such that there are no pump laser photons coherently scattered into the cavity. As we have shown, adding an extra specially tuned driving field significantly improves the performance of the system. 

At this point our simulations are limited to a few hundred atoms since we use a second-order cumulant expansion as a minimum model to reliably predict the laser spectrum. Much higher output laser power with a cleaner spectrum can be expected for realistic atom numbers of up to a million, where each atom only needs to contribute very weakly to the gain and thus less pump power is required for lasing. Unfortunately such system sizes are beyond our present numerical capabilities. Similarly, a reliable evaluation of the photon statistics, as for instance the calculation of the second-order correlation function $g^{(2)}(\tau)$, requires even higher order expansions thereby limiting the tractable atom number even further.   

Conceptually, the chosen example setup constitutes a minimalist implementation of a superradiant laser requiring only a single standing wave pump field and a single mode within the cavity to facilitate trapping, pumping, and lasing simultaneously. Via state-dependent light forces atoms excited at the antinodes of the pump standing wave are drawn towards its nodes, where the coupling to the cavity mode can be made maximal by a suitable mode choice.
The operating principle here is reminiscent of maser implementations, where in order to implement gain one uses coherent excitation and magnetic separation of the excited state fraction. In our model, the excited state separation is facilitated by differential optical gradient forces, which typically are much stronger than magnetic gradient forces for neutral atoms. 

Eventually more complex geometries involving higher-order transverse modes and special state-dependent optical guiding fields can be envisaged for better performance to increase gain and pump efficiency. As there is a large number of options here, we have restricted ourselves to only one generic implementation to exhibit the basic principle more clearly. Future more refined models need to be developed in collaboration with a specific experimental implementation. 

\section*{Acknowledgements}
We acknowledge funding from the L'ORÉAL Austria Fellowship "For Women in Science" 2023 and the European Union’s Horizon 2020 research and innovation program under the Marie Sklodowska-Curie Grant Agreement No.~860579 (project MoSaiQC).

\section*{Data Availability}
All plots were generated directly from the formulas within the paper using the Julia programming language and the open source framework QuantumOptics.jl \cite{kramer18}. The data are openly available at \href{https://doi.org/10.5281/zenodo.7950727}{\it{https://doi.org/10.5281/zenodo.7950727}}.

\bibliography{Bib_laser}

\end{document}